\documentclass[sigplan,nonacm]{acmart}
\makeatletter
\renewcommand\@formatdoi[1]{\ignorespaces}
\makeatother
\setcopyright{none}
\copyrightyear{2023}
\acmConference[]{}{}{}

\usepackage{enumitem}
\usepackage{mathtools}
\usepackage{amsmath}
\usepackage{empheq}
\usepackage{dcolumn}
\newcolumntype{d}[1]{D{.}{.}{#1}}

\usepackage{listings}

\lstdefinelanguage{pcpp}{%
  language     = C++,
  morekeywords = {type,then,of,match,with,null},
  keywordstyle=\ttfamily\bfseries
}

\usepackage{amsthm}

\usepackage{natbib}
\usepackage{url}

\newcommand{\defeq}{\stackrel{\text{def}}{=}}

\newdimen\zzlistingsize
\newdimen\zzlistingsizedefault
\newdimen\kwlistingsize
\zzlistingsize=\zzlistingsizedefault
\gdef\lco{black}

\newcommand{\Lstbasicstyle}{\fontsize{\zzlistingsize}{1.1\zzlistingsize}\ttfamily\color{\lco}}
\newcommand{\keywordstyle}{\fontsize{1.09\kwlistingsize}{\kwlistingsize}\normalfont\bf\color{\lco}}
\newlength{\zzlstwidth}
\newcommand{\lcm}{\color{\lco}}

\title{The best multicore-parallelization refactoring you've never heard of}
\titlenote{Our title is a riff on ``The Best Refactoring You've Never Heard Of''~\cite{bestrefactoring}, from the popular blog post and Compose 2019 talk.}
\author{Mike Rainey}
\affiliation{
  \institution{Carnegie Mellon University}            
  \streetaddress{Street1 Address1}
  \city{Pittsburgh}
  \state{PA}
  \postcode{Post-Code1}
  \country{USA}                    
}
\email{me@mike-rainey.site}          

\begin{document}

\begin{abstract}
In this short paper, we explore a new way to refactor a simple but
tricky-to-parallelize tree-traversal algorithm to harness multicore
parallelism.
Crucially, the refactoring draws from some classic techniques from
programming-languages research, such as the continuation-passing-style
transform and defunctionalization.
The algorithm we consider faces a particularly acute
granularity-control challenge, owing to the wide range of inputs it has
to deal with.
Our solution achieves efficiency from heartbeat scheduling, a recent
approach to automatic granularity control.
We present our solution in a series of individually simple refactoring
steps, starting from a high-level, recursive specification of the
algorithm.
As such, our approach may prove useful as a teaching tool, and perhaps
be used for one-off parallelizations, as the technique requires no
special compiler support.
\end{abstract}
\maketitle
\section{Challenge: traverse a pointer-based tree}

We are to write a program that traverses a given binary tree and
returns the sum of the numbers stored in the nodes, as exemplified by
the following reference code, taking care to utilize parallelism
when the input permits and to perform well in any case, even when
parallelism is limited.
\begin{lstlisting}[language=pcpp,numbers=none]
type node = {v : int, bs : node*[2]}
sum(node* n) $\rightarrow$ int { if (n == null) return 0
  return sum(n.bs[0]) + sum(n.bs[1]) + n.v }
\end{lstlisting}
We may use fork join to parallelize on nonempty input trees:
\begin{lstlisting}[language=pcpp,numbers=none]
s = new int[2]
fork2join($\lambda$ () $\Rightarrow$ {s[$\mathtt{i}$] = sum(n.bs[$\mathtt{i}$])}, ...) $\mathtt{i} \in \{0,1\}$
return s[0] + s[1] + n.v
\end{lstlisting}
There are mature implementations of fork
join (e.g.~\cite{threadingbuildingblocksmanual,schardl2017tapir})
based on work stealing~\cite{halstead84,blumofele99,chasele05}, an
algorithm well suited for irregular, input-dependent workloads like
ours.

However, suppose our program can assume nothing of its
input: it can range from balanced and large, where parallelism is
abundant, to, e.g., a chain or a small tree, where
traversal is mostly serial.
As such, possible workloads may be fine grain and irregular, making
granularity control
acute~\cite{lazy-binary-splitting,pdfs-15,OracleGuided_19}.
Also, inputs, e.g., long chains, may cause callstack
overflow~\cite{lee2010cilk-m}.
We address both problems by combining two known techniques:
(1) heartbeat scheduling~\cite{Heartbeat_17,TPAL_21} for
granularity control and (2) defunctionalize the
continuation~\cite{10.1145/800194.805852} to replace
recursion by iteration, for efficient, serial traversal.
The latter is inspired by Koppel's
presentation~\cite{bestrefactoring}, adding to his list a new
refactoring application: multicore parallelization.
We proceed via a series of refactorings, using conventional features of
C++.

\section{Refactoring for parallel traversal}
First, we replace the direct-style treatment that
allocates function-activation records on the C call\-stack with a
continuation-passing-style (CPS) one that allocates on the heap.
To support CPS, we use a lower-level interface with task
scheduling: \lstinline|new_task($f$)| takes a thunk $f$ and returns a
pointer to a new, heap-allocated task that, when executed by the
scheduler will run $f()$ to completion; \lstinline|fork($c$, $k$)|
takes a child task $c$, its continuation task $k$, registers a
dependency edge from $c$ to $k$, and marks $c$ ready to run;
\lstinline|join($j$)| marks an incoming dependency edge on the task
$j$ as resolved, and, when all of its dependencies are resolved,
schedules $j$.

\paragraph{Step 1: CPS convert the parallel algorithm}
We introduce a continuation parameter $\mathtt{k}$ and a \emph{join task}
$\mathtt{tj}$, which receives the results of the recursive calls and
passes the results to the return continuation.
%
\begin{lstlisting}[language=pcpp,numbers=none]
sum(node* n, k : int $\rightarrow$ void) $\rightarrow$ void {
  if (n == null) { k(0); return }; s = new int[2]
  tj = new_task($\lambda$ () $\Rightarrow$ k(s[0] + s[1] + n.v))
  { $\mathtt{t_i}$ = new_task($\lambda$ () $\Rightarrow$
       sum(n.bs[$\mathtt{i}$], $\lambda$ $\mathtt{s_i}$ $\Rightarrow$ {s[$\mathtt{i}$] = $\mathtt{s_i}$; join(tj)})
    fork($\mathtt{t_i}$, tj) } $\mathtt{i} \in \{0,1\}$ }
\end{lstlisting}

\paragraph{Step 2: defunctionalize the continuation}
We introduce one activation record to handle the final
result, and another for completion of branch $\mathtt{i} \in
\{0,1\}$ (full code in appendix).
\begin{lstlisting}[language=pcpp,numbers=none]
type kont = | KTerm of int* // final result
  | KPBranch of {i : int, s : int*, tj : task*}
\end{lstlisting}
This refactoring delivers a highly parallel algorithm, but one with
poor work efficiency, given that it performs little useful
work per task.

\section{Refactoring for serial traversal}
Now, we obtain a work-efficient version by replacing recursion with
iteration.

\paragraph{Step 3: CPS convert \& defunctionalize the continuations}
We CPS convert our reference algorithm, first by
introducing two new continuations, and then defunctionalize them, giving us
two new activation records, such that the first represents an
in-flight recursive call for the first branch of a node, and the
second for the second branch, with $\mathtt{s_0}$ storing result
obtained for the first branch.
\begin{lstlisting}[language=pcpp,numbers=none]
type kont = $\ldots$ | $\mathtt{KSBranch_0}$ of {n : node*, k : kont*}
  | $\mathtt{KSBranch_1}$ of {$\mathtt{s_0}$ : int, n : node*, k : kont*}
\end{lstlisting}

\paragraph{Step 4: refactor for iterative, stack-based traversal}
We eliminate recursion by applying to the \lstinline|apply|
function (introduced in Step 3) both tail-call elimination and
inlining, and tail-call elimination to our defunctionalized
\lstinline|sum| function.

\section{Merging parallel and serial refactorings}
%
The conceptual glue for merging our serial and parallel algorithms is
in heartbeat scheduling.
With it, we make it so that our serial and parallel traversals alternate
on a regular basis.
Starting out, our program spends a certain amount of its time in
serial traversal, specified by a heartbeat-rate parameter $H$, after which
it switches momentarily to parallel traversal.
It then switches back to serial, and the alternation repeats until the
traversal completes.

By ensuring $H$ serial traversal steps happen for each invocation of
our parallel traversal, we amortize task-creation costs, and
therefore, achieve granularity control for \emph{all} inputs.
In our implementation, we found that it suffices to specify $H$ to be
the number of trips around the main loop of our serial traversal.
On our test machine, we observed that, by experimenting with different
settings of $H$, we can bound task-creation costs such that the total
amount of work is increased by a desired amount, e.g., 10\%, compared
to the serial refactoring.

\paragraph{Step 5: give the serial traversal a heartbeat}
To track the number of steps, we introduce a helper function
\lstinline|heartbeat| that returns \lstinline|true| every $H$ times
it is called.
When it returns \lstinline|true|, we inspect the
current continuation to see if it is holding onto any latent
parallelism.
If so, we \emph{promote} that latent parallelism into an actual task,
which may realize actual parallelism (if, e.g., the task is stolen).

\paragraph{Step 6: implement promotion}
Promotion is initiated by calling \lstinline|try_promote($k$)|, which
looks for latent parallelism in $k$ and, if present, spawns from it a
task and returns a modified continuation $k'$.
There is latent parallelism in $k$ if there is an instance of
$\mathtt{KSBranch_0}$ in $k$.
The reason is that such an instance represents a recursive call to the
first branch of some tree node (the only opportunity
parallelism in a traversal).

However, there may be multiple instances of latent parallelism in a
given $k$, and, for performance reasons, heartbeat scheduling requires
that the \emph{outermost} instance is the one that should be targeted
for promotion.
Heartbeat scheduling targets outermost parallelism because doing so
turns out to be crucial for achieving worst-case bounds on the loss of
parallelism~\cite{Heartbeat_17}.
Implementing this behavior efficiently requires some care, as a na\"ive
implementation could repeatedly traverse the whole stack, leading to
quadratic blowup.
Fortunately, the blowup can be remedied by extending the continuation
structure with a double-ended list, which marks promotion
potentials~\cite{Heartbeat_17,TPAL_21}.

When it finds a \lstinline|$\mathtt{KSBranch_0}${n, k=$k'$}| activation
record in $k$, our promotion handler modifies $k$ so
that, thereafter, it is as if our (defunctionalized) parallel
algorithm was invoked at that point instead of the serial version.
This behavior is achieved by (1) allocating storage for the results of the branches, \lstinline|s = new int[2]| (2) replacing our
\lstinline|$\mathtt{KSBranch_0}$| activation record with
\lstinline|KPBranch{i=0, s=s, tj=tj}|, a task-par\-allel one
(3) spawning a new task corresponding to the second branch,
i.e., \lstinline|n.bs[1]|, and giving that task the
return continuation \lstinline|KPBranch{i=1, s=s, tj=tj}|,
and (4) creating a join task \lstinline|tj| for this new fork
point, which is seeded with the continuation of our
promotion point, $k'$.
The pseudocode below gives sketch of the main loops.

\begin{lstlisting}[language=pcpp,numbers=none]
sum(node* n, k : kont*) $\rightarrow$ void {
  while (true)
    k = try_promote(k) if heartbeat() else k
    if (n == null) { $\mathtt{s_a}$ = 0 // sum accumulator
      while (true)
        k = try_promote(k) if heartbeat() else k
        match *k with // all activation records in kont
        | $\mathtt{KSBranch_0}${n=n1, k=k1} $\Rightarrow$ { $\ldots$ } | $\ldots$ | $\ldots$
    else { k = $\mathtt{KSBranch_0}${n=n, k=k}; n = n.bs[0] } }
\end{lstlisting}
\section{Performance study}

\begin{table}
  \begin{tabular}{r|d{3.2}d{3.2}d{3.2}d{3.2}}
    input & \multicolumn{1}{c}{serial (s)} & \multicolumn{1}{c}{ours} & \multicolumn{1}{c}{cilk} & \multicolumn{1}{c}{cilk+granctrl} \\ \midrule
    perfect & 0.7 & 28.4\mathrm{x} & 15.4\mathrm{x} & 34.5\mathrm{x} \\
    random & 0.8 & 31.8\mathrm{x} & 15.3\mathrm{x} & 33.7\mathrm{x} \\
    chains & 2.5 & 11.5\mathrm{x} & \multicolumn{1}{c}{n/a} & \multicolumn{1}{c}{n/a} \\
    chain & 1.2 & 0.4\mathrm{x} & \multicolumn{1}{c}{n/a} & \multicolumn{1}{c}{n/a} \\
  \end{tabular}
  \caption{Performance results from an Intel Xeon system, using all 64
    cores, showing speedup over the iterative, serial algorithm, with
    four inputs: (1) \emph{perfect} is a perfect binary tree of height 27 (2)
    \emph{random} is a tree built from a series of path-copying
    insertions targeting random leaves (3) \emph{chains} is a small
    initial tree of height 20 extended with 30 paths of length 1
    million (4) \emph{chain} is a long chain.}
    \label{tbl:results}
\end{table}

Table~\ref{tbl:results} summarizes our performance study, for which we
used a C++ implementation.
From the \emph{perfect} tree, we see that our algorithm can achieve a
speedup comparable to that of OpenCilk~\cite{schardl2017tapir} with
manually tuned granularity control, and a speedup almost twice faster
than that of OpenCilk without granularity control.
For \emph{random}, our algorithm outperforms vanilla OpenCilk, but not
the granularity-controlled version.
The reason relates to the data structure we used in our C++ implementation to store the
activation records, an STL deque, which uses heap-allocated
chunks internally, whereas OpenCilk uses the callstack, which
is more efficient.
However, our algorithm supports long chains, whereas OpenCilk
crashes with stack overflow (indicated by cells with n/a).
From \emph{chains}, we see that our algorithm can obtain speedup even
when parallelism is somewhat scarce.
On \emph{chain}, our algorithm is about 2.5x slower serial.

\bibliographystyle{abbrv}
\bibliography{mike_rainey}

\begin{thebibliography}{10}

\bibitem{OracleGuided_19}
U.~A. Acar, V.~Aksenov, A.~Chargu{\'e}raud, and M.~Rainey.
\newblock Provably and practically efficient granularity control.
\newblock In {\em Proceedings of the 24th Symposium on Principles and Practice
  of Parallel Programming}, PPoPP '19, pages 214--228, New York, NY, USA, 2019.
  ACM.

\bibitem{Heartbeat_17}
U.~A. Acar, A.~Chargu{\'e}raud, A.~Guatto, M.~Rainey, and F.~Sieczkowski.
\newblock Heartbeat scheduling: Provable efficiency for nested parallelism.
\newblock In {\em 39th {ACM} {SIGPLAN} Conference on Programming Language
  Design and Implementation}, PLDI '18. ACM, 2018.

\bibitem{pdfs-15}
U.~A. Acar, A.~Chargueraud, and M.~Rainey.
\newblock A work-efficient algorithm for parallel unordered depth-first search.
\newblock In {\em ACM/IEEE Conference on High Performance Computing (SC)},
  pages 67:1--67:12, New York, NY, USA, 2015. ACM.

\bibitem{blumofele99}
R.~D. Blumofe and C.~E. Leiserson.
\newblock Scheduling multithreaded computations by work stealing.
\newblock {\em J. ACM}, 46:720--748, Sept. 1999.

\bibitem{chasele05}
D.~Chase and Y.~Lev.
\newblock Dynamic circular work-stealing deque.
\newblock In {\em SPAA '05}, pages 21--28, 2005.

\bibitem{halstead84}
R.~H. Halstead, Jr.
\newblock {Implementation of Multilisp: Lisp on a Multiprocessor}.
\newblock In {\em Proceedings of the 1984 ACM Symposium on LISP and functional
  programming}, LFP '84, pages 9--17. ACM, 1984.

\bibitem{threadingbuildingblocksmanual}
Intel.
\newblock Intel threading building blocks, 2011.
\newblock \url{https://www.threadingbuildingblocks.org/}.

\bibitem{bestrefactoring}
J.~Koppel.
\newblock The best refactoring you've never heard of, 2019.
\newblock
  \url{https://www.pathsensitive.com/2019/07/the-best-refactoring-youve-never-heard.html}.

\bibitem{lee2010cilk-m}
I.-T.~A. Lee, S.~Boyd-Wickizer, Z.~Huang, and C.~E. Leiserson.
\newblock Using memory mapping to support cactus stacks in work-stealing
  runtime systems.
\newblock In {\em Proceedings of the 19th International Conference on Parallel
  Architectures and Compilation Techniques}, PACT '10, pages 411--420, New
  York, NY, USA, 2010. ACM.

\bibitem{TPAL_21}
M.~Rainey, K.~Hale, R.~R. Newton, N.~Hardavellas, S.~Campanoni, P.~Dinda, and
  U.~A. Acar.
\newblock Task parallel assembly language for uncompromising parallelism.
\newblock In {\em Proceedings of the 42nd ACM SIGPLAN Conference on Programming
  Language Design and Implementation}, PLDI '21, New York, NY, USA, June 2021.
  ACM.

\bibitem{10.1145/800194.805852}
J.~C. Reynolds.
\newblock Definitional interpreters for higher-order programming languages.
\newblock In {\em Proceedings of the ACM Annual Conference - Volume 2}, ACM
  '72, page 717–740, New York, NY, USA, 1972. Association for Computing
  Machinery.

\bibitem{schardl2017tapir}
T.~B. Schardl, W.~S. Moses, and C.~E. Leiserson.
\newblock Tapir: Embedding fork-join parallelism into llvm's intermediate
  representation.
\newblock In {\em Proceedings of the 22nd ACM SIGPLAN Symposium on Principles
  and Practice of Parallel Programming}, pages 249--265, 2017.

\bibitem{lazy-binary-splitting}
A.~Tzannes, G.~C. Caragea, R.~Barua, and U.~Vishkin.
\newblock Lazy binary-splitting: a run-time adaptive work-stealing scheduler.
\newblock In {\em Symposium on Principles \& Practice of Parallel Programming},
  pages 179--190, 2010.

\end{thebibliography}

\clearpage
\appendix
\section{Reference code}
\label{sec:reference-code}
In this section, we present the pseudocode omitted from the main body
of the paper, starting with the parallel algororithm, then the serial
one, and finally the heartbeat algorithm.

\subsection{The parallel traversal}

\paragraph{Step 2: defunctionalize the continuation.}
In Figure~\ref{fig:step-2}, we show the code resulting from taking the
CPS converted version of our parallel algorithm and defunctionalizing
the continuation.

\begin{figure}
\hrule
\begin{lstlisting}[language=pcpp,numbers=none]
sum(node* n, k : kont*) $\rightarrow$ void {
  if (n == null) { apply(k, 0); return }
  s = new int[2]
  tj = new_task($\lambda$ () $\Rightarrow$
                  apply(k, s[0] + s[1] + n.v))
  { $\mathtt{t_i}$ = new_task($\lambda$ () $\Rightarrow$
            sum(n.bs[$\mathtt{i}$], KPBranch{i=i, s=s, tj=tj))
    fork($\mathtt{t_i}$, tj) } $\mathtt{i} \in \{0,1\}$  }

apply(kont* k, $\mathtt{s_a}$ : int) $\rightarrow$ void {
  match *k with
  | KPBranch{i, s, tj} $\Rightarrow$ {s[i] = $\mathtt{s_a}$; join(tj)}
  | KTerm ans $\Rightarrow$ {*ans = $\mathtt{s_a}$} }
\end{lstlisting}
\hrule
\caption{Step 2: defunctionalize the continuation}
\label{fig:step-2}
\end{figure}

\subsection{The serial traversal}

\paragraph{Step 3(a): CPS convert.}
Starting back from our our reference algorithm, we now begin the
process of turning that code into a iterative, stack-based one.
The algorithm shown in Figure~\ref{fig:step-3a} is the result of
converting our original recursive algorithm to continuation-passing
style.

\begin{figure}
\hrule
\begin{lstlisting}[language=pcpp,numbers=none]
sum(node* n, k : int $\rightarrow$ void) $\rightarrow$ void {
  if (n == null) { k(0); return }
  sum(n.bs[0], $\lambda$ $\mathtt{s_0}$ $\Rightarrow$
    sum(n.bs[1], $\lambda$ $\mathtt{s_1}$ $\Rightarrow$
      k($\mathtt{s_0}$ + $\mathtt{s_1}$ + n.v))) }
\end{lstlisting}
\hrule
\caption{Step 3(a): CPS convert}
\label{fig:step-3a}
\end{figure}

\paragraph{Step 3(b): defunctionalize the continuation.}
In this step, we defunctionalize the three continuations we introduced
in the previous step, giving us the code shown in
Figure~\ref{fig:step-3b}.
The first such continuation is the final continuation,
\lstinline|KTerm|, the continuation that receives the final result of
our \lstinline|sum| function.
The second, namely \lstinline|$\mathtt{KSBranch_0}$|, is the
continuation that receives the result of the first recursive call, and
the third, namely \lstinline|$\mathtt{KSBranch_1}$|, is the
continuation that receives the final result of the \lstinline|sum|
call, that is, \lstinline|$\mathtt{s_0}$ + $\mathtt{s_1}$ + n.v|.

\begin{figure}
\hrule
\begin{lstlisting}[language=pcpp,numbers=none]
sum(node* n, k : kont*) $\rightarrow$ void {
  if (n == null) { apply(k, 0); return }
  sum(n.bs[0], $\mathtt{KSBranch_0}${n=n, k=k})

apply(kont* k, $\mathtt{s_a}$ : int) $\rightarrow$ void {
  match *k with
  | $\mathtt{KSBranch_0}${n, k=k1} $\Rightarrow$ {
    sum(n.bs[1], $\mathtt{KSBranch_1}${$\mathtt{s_0}$=$\mathtt{s_a}$, n=n, k=k1}) }
  | $\mathtt{KSBranch_1}${$\mathtt{s_0}$, n, k=k1} $\Rightarrow$ {
    apply(k1, $\mathtt{s_0}$ + $\mathtt{s_a}$ + n.v) }
  | KTerm ans $\Rightarrow$ {*ans = $\mathtt{s_a}$} }
\end{lstlisting}
\hrule
\caption{Step 3(b): defunctionalize the continuation}
\label{fig:step-3b}
\end{figure}

\paragraph{Step 4(a): tail-call eliminate $\mathtt{apply}$.}
In this step, we get rid of the recursion in the $\mathtt{apply}$ function
from our previous step by turning it into a loop.
We do so by applying tail-call elimination, giving us the code shown
in Figure~\ref{fig:step-4a}.

\begin{figure}
\hrule
\begin{lstlisting}[language=pcpp,numbers=none]
apply(kont* k, $\mathtt{s_a}$ : int) $\rightarrow$ void {
  while (true)
    match *k with
    | $\mathtt{KSBranch_0}${n, k=k1} $\Rightarrow$ {
      sum(n.bs[1], $\mathtt{KSBranch_1}${$\mathtt{s_0}$=$\mathtt{s_a}$, n=n, k=k1})
      return }
    | $\mathtt{KSBranch_1}${$\mathtt{s_0}$, n, k=k1} $\Rightarrow$ {
      $\mathtt{s_a}$ = $\mathtt{s_0}$ + $\mathtt{s_a}$ + n.v; k = k1 }
    | KTerm ans $\Rightarrow$ {*ans = $\mathtt{s_a}$; return } }
\end{lstlisting}
\hrule
\caption{Step 4(a): tail-call eliminate $\mathtt{apply}$}
\label{fig:step-4a}
\end{figure}

\paragraph{Step 4(b): inline $\mathtt{apply}$.}
Now, we inline the $\mathtt{apply}$ function from the previous step
into the body of our \lstinline|sum| function, giving us the code in
Figure~\ref{fig:step-4b}.

\begin{figure}
\hrule
\begin{lstlisting}[language=pcpp,numbers=none]
sum(node* n, k : kont*) $\rightarrow$ void {
  if (n == null)
    while (true)
      $\mathtt{s_a}$ = 0
      match *k with
      | $\mathtt{KSBranch_0}${n, k=k1} $\Rightarrow$ {
        sum(n.bs[1], $\mathtt{KSBranch_1}${$\mathtt{s_0}$=$\mathtt{s_a}$, n=n, k=k1})
        return }
      | $\mathtt{KSBranch_1}${$\mathtt{s_0}$, n, k=k1} $\Rightarrow$ {
        $\mathtt{s_a}$ = $\mathtt{s_0}$ + $\mathtt{s_a}$ + n.v; k = k1 }
     | KTerm ans $\Rightarrow$ {*ans = $\mathtt{s_a}$; return }
    return
  sum(n.bs[0], $\mathtt{KSBranch_0}${n=n, k=k})
\end{lstlisting}
\hrule
\caption{Step 4(b): inline $\mathtt{apply}$}
\label{fig:step-4b}
\end{figure}

\paragraph{Step 4(c): tail-call eliminate $\mathtt{sum}$}
Here, we get rid of the recursion in the \lstinline|sum| function from our
previous step.
To this end, we apply tail-call elimination, giving us the code shown
in Figure~\ref{fig:step-4c}.

\begin{figure}
\hrule
\begin{lstlisting}[language=pcpp,numbers=none]
sum(node* n, k : kont*) $\rightarrow$ void {
  while (true)
    if (n == null)
      while (true)
        $\mathtt{s_a}$ = 0
        match *k with
        | $\mathtt{KSBranch_0}${n=n1, k=k1} $\Rightarrow$ {
          n = n1.bs[1]
          k = $\mathtt{KSBranch_1}${$\mathtt{s_0}$=$\mathtt{s_a}$, n=n1, k=k1}
          break }
        | $\mathtt{KSBranch_1}${$\mathtt{s_0}$, n, k=k1} $\Rightarrow$ {
          $\mathtt{s_a}$ = $\mathtt{s_0}$ + $\mathtt{s_a}$ + n.v; k = k1 }
        | KTerm ans $\Rightarrow$ {*ans = $\mathtt{s_a}$; return }
    else
      k = $\mathtt{KSBranch_0}${n=n, k=k}
      n = n.bs[0] }
\end{lstlisting}
\hrule
\caption{Step 4(c): tail-call eliminate $\mathtt{sum}$}
\label{fig:step-4c}
\end{figure}

\subsection{The heartbeat traversal}

\paragraph{Step 5: give the traversal a heartbeat.}
Now, we are going to merge the final versions of the parallel and
serial algorithms we obtained in the previous steps.
The result of our merging is shown in Figure~\ref{fig:step-5}.
In particular, it is the merging of our defunctionalized parallel
algorithm from Figure \ref{fig:step-2} and the serial algorithm in
Figure~\ref{fig:step-4c}.
For our merging, we introduce two calls to \lstinline|try_promote|,
which have the effect of controlling the switching between serial and
parallel traversals.
The switching is enabled by simply merging the \lstinline|match|
expressions of the serial and parallel versions.

\begin{figure}
\hrule
\begin{lstlisting}[language=pcpp,numbers=none]
sum(node* n, k : kont*) $\rightarrow$ void {
  while (true)
    k = try_promote(k) if heartbeat() else k
    if (n == null)
      $\mathtt{s_a}$ = 0
      while (true)
        k = try_promote(k) if heartbeat() else k
        match *k with
        | $\mathtt{KSBranch_0}${n=$\mathtt{n_1}$, k=k1} $\Rightarrow$ {
          n = $\mathtt{n_1}$.bs[1]
          k = $\mathtt{KSBranch_1}${$\mathtt{s_0}$=sa, n=$\mathtt{n_1}$, k=k1}
          break }
        | $\mathtt{KSBranch_1}${$\mathtt{s_0}$, n=$\mathtt{n_1}$, k=k1} $\Rightarrow$ {
          s += $\mathtt{s_0}$ + $\mathtt{n_1}$.v; k = k1 }
        | KPBranch {i, s, tj} $\Rightarrow$ {
          s[i] = sa; join(tj); return }
        | KTerm ans $\Rightarrow$ { *ans = sa }
    else { k = $\mathtt{KSBranch_0}$ {n=n, k=k}; n = n.bs[0] } }
\end{lstlisting}
\hrule
\caption{Step 5: give the traversal a heartbeat.}
\label{fig:step-5}
\end{figure}

\paragraph{Step 6: implement promotion.}
All that remains of our implementation is the code that handles
promotions, which we show in Figure~\ref{fig:step-6}.
In this function, we search in our input continuation \lstinline|k|
for an instance of the activation record \lstinline|$\mathtt{KSBranch_0}$|, which
represents the continuation waiting for the result of the first branch.
If this search fails, then it returns a \lstinline|null| pointer value
in \lstinline|kt|, and we exit early by returning the original
continuation \lstinline|k|.
Otherwise, a sucessful search means that we have captured
an instant in our serial traversal when it is in the middle of traversing
the first branch of a tree node.
For such a case, we can parallelize the in-flight traversal of that
first branch with the yet-to-start traversal of the corresponding right
branch.

\begin{figure}
\hrule
\begin{lstlisting}[language=pcpp,numbers=none]
try_promote(k : kont*) $\rightarrow$ kont* {
  kt = find_outermost(k, $\lambda$ k $\Rightarrow$ {
                        match *k with
                        | $\mathtt{KSBranch_0}$ _ $\Rightarrow$ true
                        | _ $\Rightarrow$ false })
  if (kt == null) { return k }
  match *kt with
  | $\mathtt{KSBranch_0}${n, k=kj} $\Rightarrow$ {
    s = new int[2]
    t1 = new_task($\lambda$ () $\Rightarrow$ {
           sum(n.bs[1], KPBranch {i=1, s=s, tj=tj})})
    tj = new_task($\lambda$ () $\Rightarrow$ {
           $\mathtt{k_0}$ = $\mathtt{KSBranch_1}${$\mathtt{s_0}$=s[0] + s[1], n=n, k=kj}
           sum(null, $\mathtt{k_0}$) })
    fork(t1, tj)
    $\mathtt{k_1}$ = KPBranch {i=0, s=s, tj=tj}
    return replace(k, kt, $\mathtt{k_1}$) } }

// returns a pointer value $\mathtt{k1}$ to the outermost activation record in $\mathtt{k}$
// s.t. $\mathtt{f}$($\mathtt{k1}$), or null if there is no such $\mathtt{k1}$ in $\mathtt{k}$
find_outermost(k : kont*, f : kont* $\rightarrow$ bool) $\rightarrow$ kont*

// returns a pointer value $\mathtt{k1}$ s.t. any frame $\mathtt{kt}$ in
// $\mathtt{k}$ is replaced by $\mathtt{k_1}$
replace(k : kont*, kt : kont*, $\mathtt{k_1}$ : kont*)
  $\rightarrow$ kont*
\end{lstlisting}
\hrule
\caption{Step 6: implement promotion}
\label{fig:step-6}
\end{figure}

The rest of this function spawns a new task for the right branch following
the pattern in the code shown in Figure~\ref{fig:step-2}.
It uses the \lstinline|replace| function to rewrite the input continuation
in place such that the \lstinline|$\mathtt{KSBranch_0}$| activation
record is replaced by a \lstinline|KPBranch| activation record.
The effect of this step is to change the behavior of the in-flight
traversal of the affected branch so that it synchronizes with its
sibling task once its part of the traversal finishes.

\paragraph{Optimizing the layout of continuation records}
One final optimization needed for work efficiency relates to
the representation of the continuation.
As it is currently, continuation records are laid out in different
heap objects, and are linked by explicit \lstinline|kont*| pointers.
This representation can be improved by simply laying out these records
in a linear fashion, which we do by packing the records in a STL deque
container.
This way, our final implementation is more compact, as it allows us to
do without explicit tail-pointer values.

\section{CPS conversion of \texttt{fork2join}}
\label{sec:cps-fork2join}
Originally, we introduced two ways of achieving fork-join parallelism:
the direct style as in our \lstinline|fork2join| primitive and the
CPS-friendly library interface.
Here, we show the direct connection between the two versions in terms
of a CPS transformation, as shown in Figure~\ref{fig:f2j}.

\begin{figure}
\begin{align*}
  E \left[ \mathtt{fork2join}\left(f_0, f_1\right) \right]_k &\defeq \{ \\
  & \mathtt{tj} \; \mathtt{=} \; \mathtt{new\_task}\left(k\right) \\
  & \mathtt{t}_0 \; \mathtt{=} \; \mathtt{new\_task}(E \left[ f_0\left(\right) \right]_{\lambda \; \mathtt{()} \; . \; \mathtt{join(tj)}} ) \\
  & \mathtt{t}_1 \; \mathtt{=} \; \mathtt{new\_task}(E \left[ f_1\left(\right) \right]_{\lambda \; \mathtt{()} \; . \; \mathtt{join(tj)}} ) \\
  & \mathtt{fork}\left(\mathtt{t}_0, \mathtt{tj} \right)\mathtt{;} \; \mathtt{fork}\left(\mathtt{t}_1, \mathtt{tj} \right) \; \} 
\end{align*}
\caption{CPS conversion rule for \lstinline|fork2join|}
\label{fig:f2j}
\end{figure}

\end{document}